\documentclass[a4paper,onecolumn]{agn6}
\usepackage{natbib}
\usepackage{graphicx}
\usepackage[a4paper]{hyperref}
\idline{}{1}
\begin{document}
   \title{Nuclear Starbursts in Nearby Seyfert Galaxies
}

   \author{A. Romano \inst{1,2}, 
   A. Rifatto \inst{2},
           S.Ciroi \inst{1},
          P. Rafanelli \inst{1},
M. Dopita \inst{3}
}

   \institute{Dipartimento di Astronomia, Universit\`a di Padova,
vicolo dell'Osservatorio 2, 35122 Padova, Italy 
         \and
             Osservatorio Astronomico di Capodimonte, via
Moiariello 16, 80131 Napoli, Italy
      \and      
Research School of Astronomy and Astrophysics, Australian National University, Private Bag, 
Weston Creek Post Office, ACT 2611, Australia }

   \abstract{Active galaxies are often associated with starbursts, which are possible 
sources of fuelling for the nuclear black holes. By means of optical 
($3500-7500$ \AA) spectroscopic and narrow-band photometric data (H$\alpha$ + 
[NII]$\lambda$6548,6583 and [OIII]$\lambda$5007) it is possible to detect starburst 
features in active nuclei and to study the correlation with Seyfert type and the 
surrounding environment.
Here we present evidences of young stellar populations (HI high order Balmer 
lines) and extended H$\alpha$ emission (indicative of HII regions) in 6 nearby 
Seyfert galaxies ($z < 0.02$). These very preliminary results seem to suggest 
that  the nuclear starburst presence does not depend on the Seyfert type, 
favouring the hypothesis of the AGN Unified Model. By estimating the present 
SFR ($< 10^6$ years) and recent SFR ($< 10^9$ years) we observe that these objects 
have experienced most of star formation episodes in past epochs. This supports 
the hypothesis of a starburst-AGN connection, favouring an evolutionary 
scenario in which the AGN survives to the starburst.
   }
   \authorrunning{A. Romano et al.}
   \titlerunning{}
   \maketitle
%
%________________________________________________________________

\section{Introduction}

One of the most important questions concerning the Active Galactic Nuclei (AGNs) 
phenomenon is the connection between starburst and nuclear activity. The role of 
starbursts in AGNs has been extensively discussed in the past and several 
theoretical scenarios have been proposed, from those in which the two phenomena 
are indirectly connected because both are triggered by and live on gas fuelling 
(e.g., mergers, interactions between galaxies, bars, etc.) to those in which the
starbursts are directly bound with the AGNs as possible sources of fuelling for 
the nuclear black holes. 
The study of the extended circum- and extra-nuclear regions of ionized gas in 
active galaxies can shed more light on the interaction between the active nucleus 
and its environment, because the gas is a convenient tracer of massive 
star-formation (HII regions), ionization by non thermal processes (e.g., 
power-law continua), and/or shocks. 
Evidences of starburst features have been detected in numerous Seyfert 2 \citep{cid}, 
but rarely in Seyfert 1 galaxies \citep{rodriguez}.
On the contrary, if the AGNs Unified Model \citep{antonucci} is valid and a connection exists 
between AGN and starburst, one would expect to observe a similar distribution of circumnuclear star-forming
regions among the different types of AGNs.
In this poster we present preliminary results of an investigation of the nuclear 
and circumnuclear regions of 6 nearby Seyfert galaxies. Narrow-band images 
isolating the emission lines of H$\alpha$+ [NII]$\lambda$6548,6583 and [OIII]$\lambda$5007 
were used to search for spatially extended circumnuclear emission regions. 
These observations were followed up with long-slit, medium resolution spectrophotometry to probe 
further the ionization state of the gas. The studied galaxies involve all Seyfert 
types (three Seyfert 2, one Seyfert 1, one Seyfert 1.5 and one Seyfert 1.8) in order to 
check a possible correlation between the presence of starburst and the Seyfert type.

%                                              large figure (place early!)
%______________________________________________ Gamma_1 (lg rho, lg e)
   \begin{figure}
   \centering
   \resizebox{10.0cm}{!}{\includegraphics{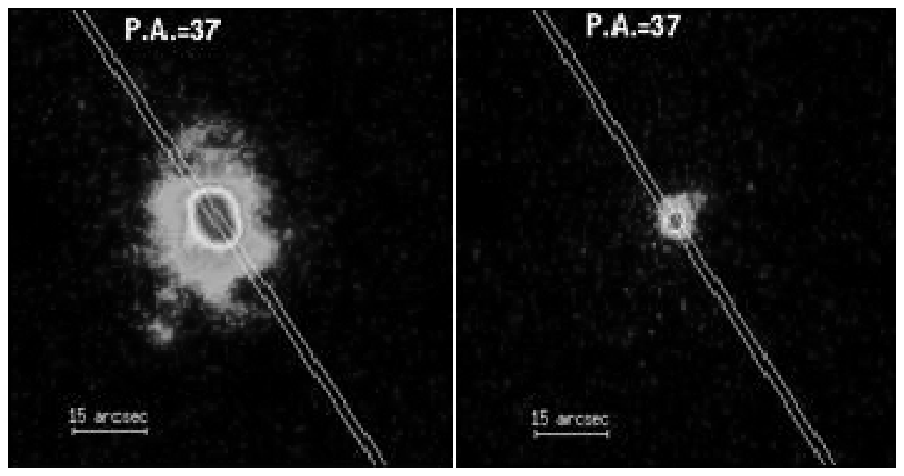}}
\vskip 20pt
   \resizebox{10.0cm}{!}{\includegraphics{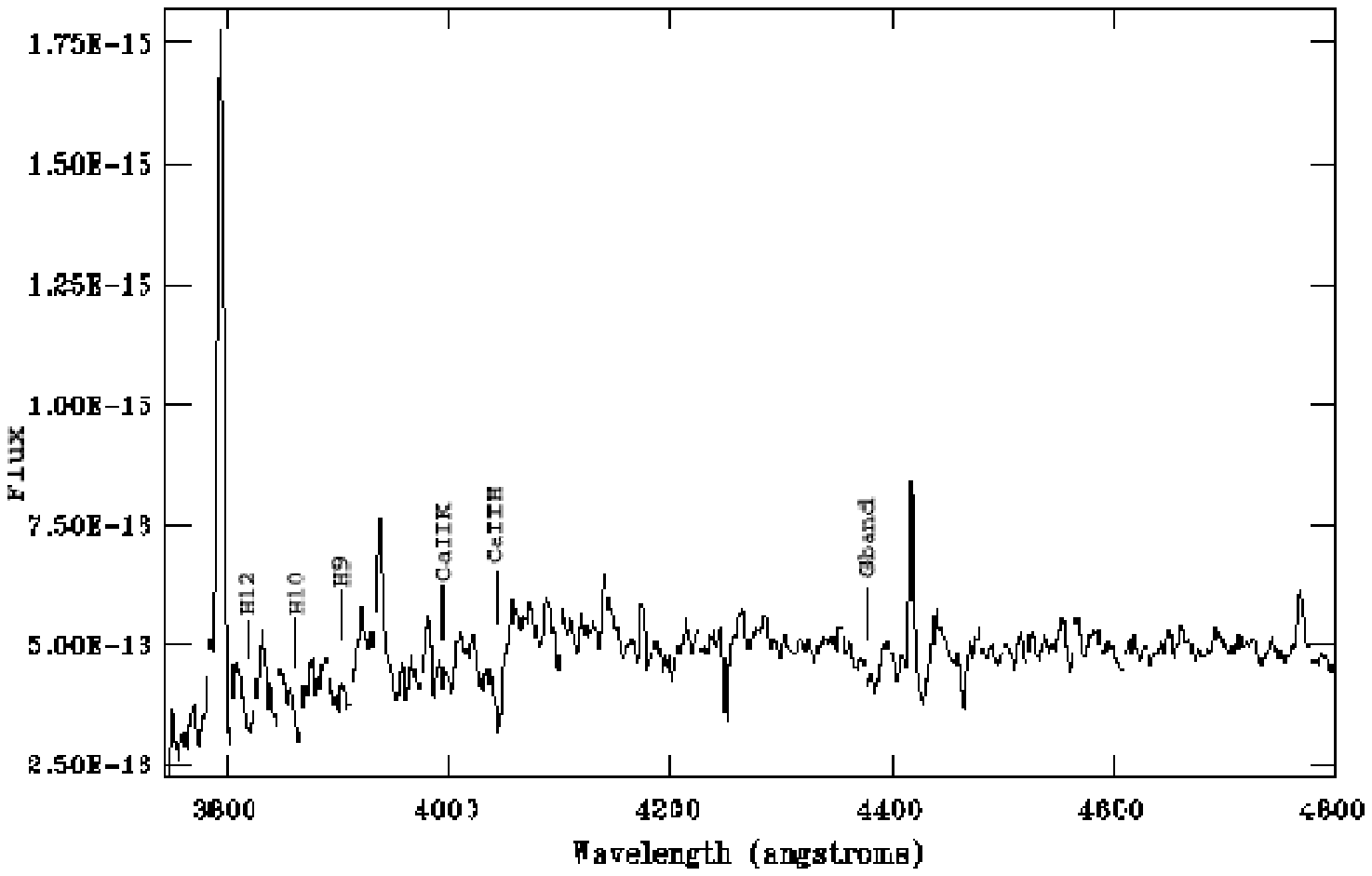}} 
   \caption{Upper panel: The $H\alpha$ (left) and $[OIII]\lambda 5007$ (right) emission-line
images of ESO018-G09. Bottom panel: Section of the nuclear spectrum of ESO018-G09 (Seyfert 2), where we see
the absorption lines of young stars.}
              \label{Fig1}
    \end{figure}
%______________________________________________________________

%
%                                                Figure with two panels
%----------------------------------------------------------- S_vib   
  \begin{figure}
  \centering
  \resizebox{10.0cm}{!}{\includegraphics{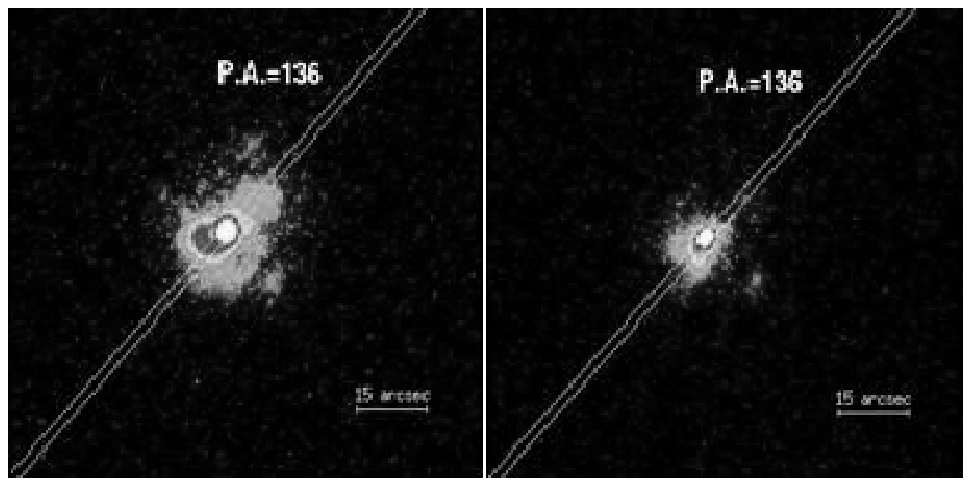}}
\vskip 20pt
   \resizebox{10.0cm}{!}{\includegraphics{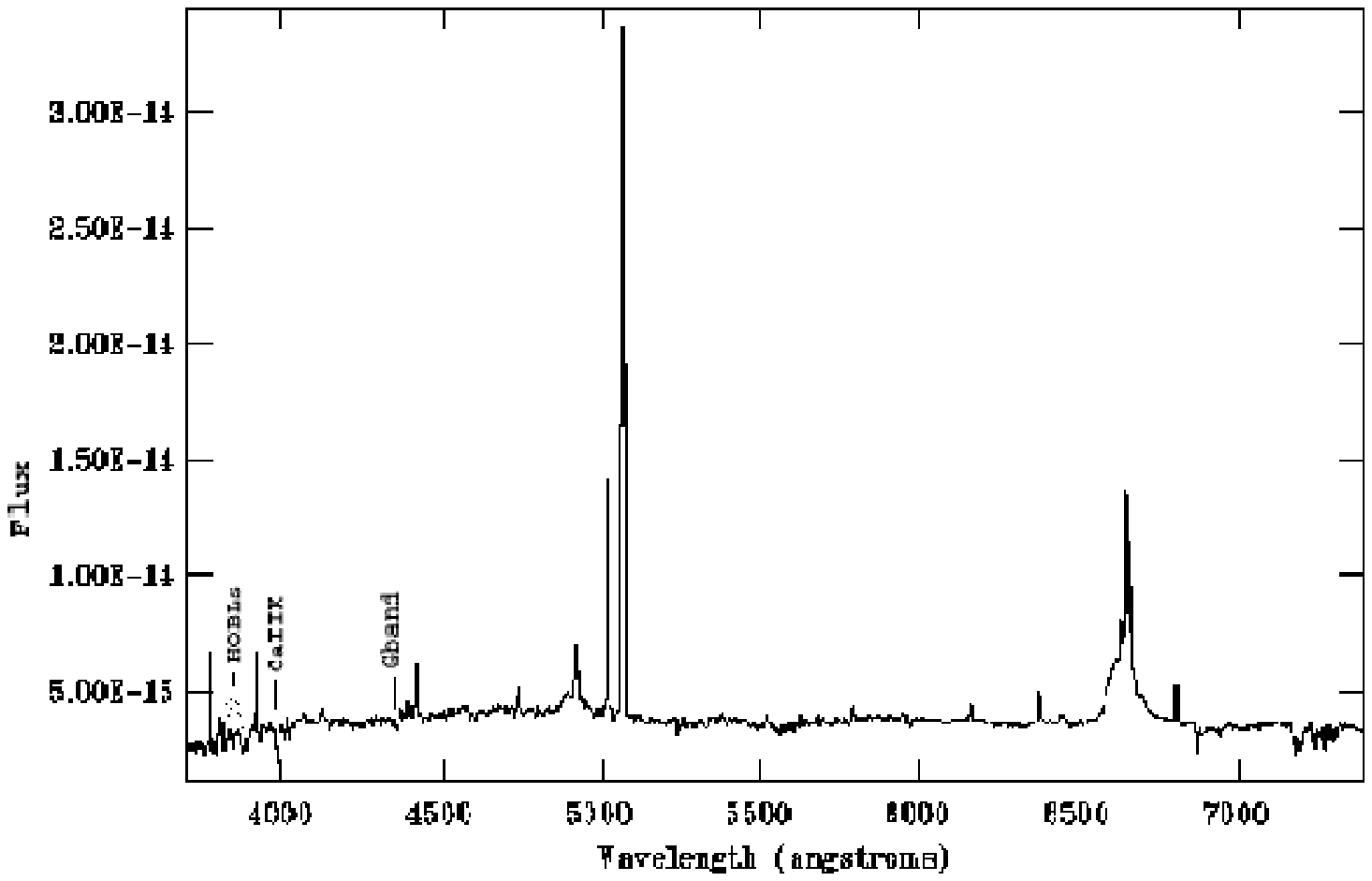}} 
   \caption{Upper panel: The $H\alpha$ (left) and $[OIII]\lambda 5007$ (right) emission-line 
images of ESO362-G18 (Seyfert 1.8). Bottom panel: The nuclear spectrum of ESO362-G18, where the absorption lines
of young stars are detectable.}
       \label{fit1}
    \end{figure}
%
%______________________________________________________________

\section {The observations}

The photometric observations were carried out in October 2001 and January 2002 at the 
Siding Spring Observatory (Australia) with 2.3m telescope. A 1kx1k CCD camera (pixel 
size = 24 $\mu m$, scale = 0.6\arcsec/px) was used in combination with H$\alpha$+ [NII]$\lambda$6548,6583 and 
[OIII]$\lambda$5007 filters (FWHM $\sim 75$ \AA). Long-slit spectra of 4 out of 6 galaxies (ESO018-G09,
 ESO362-G18, ESO377-G24, ESO428-G14) were obtained at the same telescope with the Dual Beam
 Spectrograph (pixel size = 15 $\mu m$, scale = 0.89\arcsec/px), which allowed to obtain simultaneously
 the blue (3640-5550 \AA) and the red (5540-7450 \AA) parts of the optical wavelength range at 
medium resolution ($\sim 2$ \AA). The spectra of the other 2 galaxies were obtained in December 1990
 and March 1991 at ESO-La Silla (Chile) at the 1.52m telescope with the B\&C Spectrograph 
(pixel size = 15 $\mu m$, scale = 0.82\arcsec/px) at lower resolution. The covered spectral range was 
4620-7470 \AA\ for NGC 3081 and 3730-6780 \AA\ for NGC 1365.
The data were reduced and analyzed with the IRAF software. All the observations were bias 
subtracted, flat field corrected, cleaned from cosmic rays and sky subtracted. 
The emission-line images were calibrated in flux using a planetary nebula observed in the 
same runs with the same instrumental configuration. The two-dimensional spectra were 
calibrated in wavelength using a Ne-Ar lamp for SSO spectra and He-Ar lamp for ESO spectra, 
then calibrated in flux using a spectrophotometric standard star.
For each galaxy we extracted spectra of the nucleus and several circum- and/or extra-nuclear
 regions, whose sizes were determined by means of both the H$\alpha$ line profile along the slit 
and the extension of the emission in the H$\alpha$ images. 
We identified the stellar absorption lines and bands in each spectrum. Then, the template of
 a S0 galaxy was subtracted to remove the underlying stellar population contribution. After 
this step, we identified the emission lines in the spectra of each object: their profiles 
were decomposed into multiple Gaussians using the IRAF task FITPROF. 
In particular, we measured the redshift, width and flux of each identified line. Then, the 
line fluxes were corrected for internal reddening using the observed Balmer decrement H$\alpha$/H$\beta$
chosen equal to 3.1 for the active nuclei and 2.85 for the starburst regions \citep{osterbrock}, and applying 
the extinction curve by \citet{cardelli}.

\section{Results}

We found clear evidences of the presence of young stellar populations (HI high order Balmer 
absorption lines (HOBLs), indicative of age $< 1$ Gyr) and extended H$\alpha$ emission (indicative of HII regions)
 in 4 galaxies of the sample: a Seyfert 2, a Seyfert 1, a Seyfert 1.5 and a Seyfert 1.8 (see figures 1-2). 
These preliminary results seem to suggest that the nuclear starburst presence does not depend on the 
Seyfert type, favouring the hypothesis of the AGN Unified Model. 
We also estimated the present SFR (on time-scale of $10^6$ years) and compared it with the recent 
SFR (on time-scale of $10^9$ years) for every object of the sample. We observed that the galaxies 
which do not show clear signs of present starbursts have had major star formation in the past 
epochs. This result supports the hypothesis of a starburst-AGN connection, favouring an 
evolutionary scenario in which the AGN survives to the starburst. In this scenario we should 
expect that the composed Seyfert+starburst system evolves into a “pure” Seyfert.
Moreover, we observed that the presence of nuclear starburst features is usually associated to 
the presence of extended star formation regions spread over the disk.
Given the small number of the studied objects, we cannot suggest any hypothesis at this stage 
about the role of the interactions in the starburst-AGN connection. But our future goals are 
just to extend this analysis to a larger sample of isolated Seyfert galaxies in order to verify 
these preliminary results on a statistically significant sample. 

\begin{acknowledgements}
This research was partially based on data from a INAF-OAC and ANU-MSSSO project, in collaboration with the 
Department of Astronomy, University of Padova.
 %     We gratefully acknowledge.. Part of this work was supported by.. 
\end{acknowledgements}

\bibliographystyle{aa}

\end{document}